\documentclass[useAMS,usenatbib,a4paper,referee]{mn2e}
\usepackage{graphicx}
\usepackage{amsmath}
\usepackage{txfonts}
\usepackage{mathrsfs}

\def\aap{A\&A}
\def\apjl{ApJ}

\def\apjs{ApJS}

\newcommand{\be}{\begin{equation}}
\newcommand{\ee}{\end{equation}}
\newcommand{\bea}{\begin{eqnarray}}
\newcommand{\eea}{\end{eqnarray}}

\title[Cosmological Redshift]{Cosmological Redshift in FRW Metrics with Constant Spacetime Curvature}
\author[Fulvio Melia]{Fulvio Melia\thanks{John Woodruff Simpson Fellow. E-mail: melia@as.arizona.edu}\\
Department of Physics, The Applied Math Program, and Department of Astronomy, 
The University of Arizona, AZ 85721, USA}

\begin{document}

\date{}

\pagerange{\pageref{firstpage}--\pageref{lastpage}} \pubyear{2010}

\maketitle

\label{firstpage}

\begin{abstract}
Cosmological redshift $z$ grows as the Universe expands and is conventionally viewed
as a third form of redshift, beyond the more traditional Doppler and gravitational
effects seen in other applications of general relativity. In this paper, we examine the
origin of redshift in the Friedmann-Robertson-Walker metrics with constant spacetime
curvature, and show that---at least for the static spacetimes---the interpretation of 
$z$ as due to the ``stretching" of space is coordinate dependent. Namely, we prove that
redshift may also be calculated solely from the effects of kinematics and
gravitational acceleration. This suggests that its dependence on the expansion
factor is simply a manifestation of the high degree of symmetry in FRW, and 
ought not be viewed as evidence in support of the idea that space itself is 
expanding.
\end{abstract}

\begin{keywords}
{cosmic microwave background, cosmological parameters, cosmology: observations,
cosmology: redshift, cosmology: theory, distance scale}
\end{keywords}

\section{Introduction}
Standard cosmology is based on the Friedmann-Robertson-Walker (FRW) metric for a spatially
homogeneous and isotropic three-dimensional space. In terms of the proper time $t$ measured
by a comoving observer, and the corresponding radial ($r$) and angular ($\theta$ and $\phi$)
coordinates in the comoving frame, the interval  
\begin{equation}
ds^2=g_{\mu\nu}\,dx^{\,\mu} dx^\nu\;,
\end{equation}
where $g_{\mu\nu}$ ($\mu,\nu=0,1,2,3$) are the metric coefficients,
may be written as
\begin{equation}
ds^2 = c^2 dt^2 - a^2(t)[dr^2 (1 - kr^2)^{-1} + r^2(d\theta^2 + \sin^2\theta d\phi^2)]\;.
\end{equation}
The expansion factor $a(t)$ is a function of cosmic time $t$, whereas the spatial coordinates
$(r,\theta,\phi)$ in this frame remain ``fixed" for all particles in the cosmos. The constant 
$k$ is $+1$ for a closed universe, $0$ for a flat, open universe, or $-1$ for an open universe. 

This representation of proper distance as a product of a universal expansion factor (independent
of position) and an unchanging set of comoving coordinates, is often interpreted as meaning
that space itself is dynamic, expanding with time. This view, however, is not universally
accepted because the difference between this situation---in which particles are fixed in 
an expanding space---and the alternative interpretation---in which the particles move 
through a fixed space---is more than merely semantic. Each has its own particular set 
of consequences, some of which have been explored elsewhere, e.g., by Chodorowski (2007).

Underlying much of the discussion concerning the expansion of space (see, e.g., Davis \&
Lineweaver 2004; Harrison 1995; Chodorowski 2007; Baryshev 2008; Bunn \& Hogg
2009; Cook \& Burns 2009) is the nature of cosmological redshift $z$, defined as
\begin{equation}
z={\nu_e-\nu_o\over\nu_o}\;,
\end{equation}
where $\nu_o$ and $\nu_e$ are the observed and emitted radiation frequencies,
respectively. It is not difficult to show (Weinberg 1972) that
\begin{equation}
1+z={a(t_o)\over a(t_e)}\;,
\end{equation}
in terms of the expansion factor $a(t)$, where $t_o$ and $t_e$ represent,
respectively, the cosmic time at which the radiation is observed and that at 
which it was emitted. It is this formulation, in particular, that seems to suggest 
that $z$ is due to the aforementioned stretching of space, because it doesn't 
look like any of the other forms of redshift we have encountered before. 
But is cosmological redshift really due to ``stretching," and therefore a 
different type of wavelength extension beyond those expected from 
Doppler and gravitational effects? Or is this different 
formulation---and therefore its interpretation---merely due to our
choice of coordinates? In other words, is it possible to use another set of
coordinates to cast the cosmological redshift into a form more like 
the ``traditional" lapse function used in other applications of general relativity? 
This is the principal question we wish to explore in this paper.

But finding a resolution to this important issue is quite difficult, as others have
already discovered (see, e.g., Bunn \& Hogg 2009; Cook \& Burns 2009). In this 
paper, we will seek a partial answer to this question by considering
a subset of FRW metrics---those that have a constant 
spacetime curvature and can therefore be written in static form. The complete
treatment, including also those FRW metrics whose curvature changes with
time, will be discussed elsewhere. For these static FRW metrics, we will
prove that the cosmological redshift can
be calculated---with equal validity---either from the ``usual" expression
(Equation~4) involving the expansion factor $a(t)$, or from the well-known
effects of kinematic and gravitational time dilation, using a transformed
set of coordinates $(cT,\eta,\theta,\phi)$, for which the metric coefficients
$g_{\mu\nu}$ are independent of time $T$. We will therefore show for the
static FRW metrics, that the interpretation of $z$ as a stretching of
space is coordinate-dependent. A different picture emerges when we 
derive $z$ directly as a ``lapse function" due to Doppler and gravitational
effects.

\section{The Cosmological Lapse Function}
Our procedure for finding the cosmological redshift as a lapse function
involves three essential steps. First, we find a set of coordinates permitting 
us to write the metric in stationary form. It goes without saying that
Equation~(2) is not adequate for our purposes because the 
metric coefficients $g_{\mu\nu}$ generally depend on time $t$,
through the expansion factor $a(t)$. Second, we use this transformed 
metric to calculate the time dilation at the emitter's location relative to
the proper time in a local free-falling frame. Finally, we obtain
the {\it apparent} time dilation, which differs from its counterpart
at the emitter's location because the motion of the source alters the
relative arrival times of the photon's wave crests. Steps two and three
are rather standard in relativity (see, e.g., Weinberg 1972). The most complicated
portion of this procedure is the search for an appropriate coordinate
transformation that renders the FRW metric static.

It is not difficult to show that there are exactly six FRW metrics with constant spacetime
curvature; in each of these cases, a transformation of coordinates permits
us to write these solutions in static form (Florides 1980). We will consider each of
these special cases in turn, including the Minkowski spacetime, the Milne Universe,
de Sitter space, the Lanczos Universe, and anti-de Sitter space.  It is important to stress as we proceed 
through this exercise that although the spacetime curvature is constant in the cases
we consider here, it is generally nonzero. This is a crucial point because the cosmological 
redshift is therefore not just a kinematic effect (as in the Milne Universe); it is generally 
a combination of Doppler and gravitational effects (as one finds in de Sitter and 
Lanczos). Static FRW metrics therefore do not simplify the redshift by eliminating one 
or more of the contributors.  Gravitational effects are present even when the FRW 
metric is time-independent, as is well known from the Schwarzschild and Kerr
spacetimes.
\vfill
\section{The Six Static FRW Metrics}

\subsection{Minkowski Spacetime}
The Minkowski spacetime is spatially flat ($k=0$) and is not expanding,
$a(t)=1$, so
\begin{equation}
ds^2=c^2dt^2-dr^2-r^2\,d\Omega^2\;,
\end{equation}
where, for simplicity, we have introduced the notation
$d\Omega^2\equiv d\theta^2+\sin^2\theta\,d\phi^2$.
This metric is already in static form, so there is no need to find a new
set of coordinates. Quite trivially, then, $z=0$ everywhere (from Equation~4). 
The Doppler and gravitational redshifts are also trivially zero in this case, since 
there is no expansion or spacetime curvature. 

\subsection{The Milne Universe}
A universe with $\rho=0$ and $k=-1$ corresponds to a simple solution of
Einstein's equations, in which
\begin{equation}
a(t)=ct\;,
\end{equation}
i.e., the scale factor grows linearly in time. Since the ``acceleration" $\ddot{a}(t)$ is
therefore zero in this cosmology, first introduced by Milne (1933), one
might expect such a universe to be flat and a mere re-parametrization of
Minkowski space. Indeed, Milne intended this type of expansion to be
informed only by special relativity, without any constraints imposed by
the more general theory. This cosmology has been the subject of many
past analyses, including two recent publications (Abramowicz et al.
2007; Cook \& Burns 2009)
that considered it in its manifestly flat form, obtained through a
straightforward coordinate transformation that we describe as follows.

We first introduce the co-moving distance variable $\chi$, defined in
terms of $r$ according to
\begin{equation}
r=\sinh\chi\;,
\end{equation}
which allows us to write the FRW metric for Milne in the form
\begin{equation}
ds^2=c^2dt^2-c^2t^2[d\chi^2+\sinh^2\chi\,d\Omega^2]\;.
\end{equation}
The transformation that brings Equation~(8) into a stationary
(and manifestly flat) form is
\begin{eqnarray}
\hskip 2.7in T&\hskip-0.1in=\hskip-0.1in&t\cosh\chi\nonumber\\
\eta&\hskip-0.1in=\hskip-0.1in&ct\sinh\chi\;,
\end{eqnarray}
for then
\begin{equation}
ds^2=c^2dT^2-d\eta^2-\eta^2d\Omega^2\;.
\end{equation}
The fact that this form of the metric is identical to the Minkowski spacetime 
(Equation~5), confirms the generally understood identity between the two
or, as we alluded to above, the fact that one is a re-parametrization of
the other. Neither Minkowski nor Milne have any spacetime curvature,
and may therefore be transformed into each other with an appropriate
set of coordinates (as we have just seen). However, we will affirm on several 
occasions in the following sections that the measurement of redshift depends 
critically on the observer and the coordinates he/she is using. Thus, even
though Minkowski and Milne are equivalent, the source is moving (with the
Hubble flow) relative to an observer in the latter, but not the former, and the 
two observers therefore do measure a different kinematic redshift.   

For the second step, let us now evaluate the time dilation in
the coordinate frame $(cT,\eta,\theta,\phi)$, assuming only
radial motion, i.e., $d\theta=d\phi=0$. In the co-moving
frame, the cosmic time $t$ is also the proper time (what we would
conventionally call $\tau$). Thus, for an interval
associated with proper time only, (i.e., $ds^2=c^2dt^2$),
we have from Equation~(10)
\begin{equation}
{dt\over dT}=\left[1-{1\over c^2}\left({d\eta\over dT}\right)^2\right]^{1/2}\;.
\end{equation}

This time dilation, however, evaluated at the emitter's location (and at
the time when the light was produced), is not necessarily equal to the
{\it apparent} time dilation. These two are equal only when the source
is instantaneously at rest with respect to the observer. If the source
is moving (as it is here), then the time between emission of successive
wave fronts is indeed given by $dT$ in Equation~(11), but during this
interval, the proper distance (as measured in the $\eta-T$ frame)
from the observer to the light source also
increases by an amount $v_\eta\sqrt{g_{TT}}\,dT$, where
\begin{equation}
v_\eta\equiv \sqrt{g_{\eta\eta}\over g_{TT}}{d\eta\over dT}
\end{equation}
is the component of (proper) velocity (proper distance per unit proper time)
measured in this frame along the line-of-sight to the source.

Thus, the ratio of the frequency of light actually measured by the observer
to that emitted is
\begin{equation}
{\nu_o\over\nu_e}=\left(1+{v_\eta\over c}\right)^{-1}{dt\over dT}
\,{\bigg{|}}_{T_e}\;,
\end{equation}
a simple expression made possible by the {\it static} form of the
metric in Equation~(10). If the metric coefficients $g_{\mu\nu}$
had been dependent on $T$, other multiplicative factors would
have needed to be introduced into Equation~(13). (Note that the quantities 
on the right-hand side of this equation formally must all be evaluated at the 
time, $T_e$ or, equivalently, $t_e$, when the light was emitted. In the Milne 
Universe, the expansion velocity at a fixed $\chi$ is trivially constant in time.
This criterion is much more important for the curved spacetimes we will
consider next.)

Assuming that the source is moving with the Hubble flow, i.e., that
$dr=0$, we now see that
\begin{equation}
{d\eta\over dT}={\partial\eta\over \partial t}{dt\over dT}=c\tanh\chi
\end{equation}
and Equations~(11) and (14) are therefore trivially consistent with
\begin{equation}
{dt\over dT}={1\over\cosh\chi}\;.
\end{equation}
Since in the Milne cosmology $g_{\eta\eta}=g_{TT}=1$, the apparent frequency
shift is therefore
\begin{eqnarray}
\hskip2.2in{\nu_o\over\nu_e}&=&(1+\tanh\chi)^{-1}\cosh^{-1}\chi\nonumber \\
&=&e^{-\chi}\;.
\end{eqnarray}
So according to this procedure for finding $z$ via the lapse function,
the cosmological redshift is given by 
\begin{equation}
1+z\equiv {\nu_e\over \nu_o}=e^\chi\;.
\end{equation}
How does this compare with the expression one would conventionally
derive from Equation~(4), based on the rate of universal expansion
between the emission ($t_e$) and observation ($t_o$) times?

In starting its propagation from the source at time $t_e$, the emitted
light travels along a null geodesic ($ds=0$) until it reaches the observer
at time $t_0$. Therefore, from Equation~(8) with $d\Omega=0$, we see that
\begin{equation}
\int_0^\chi d\chi' = \int_{t_e}^{t_o}{dt'\over t'}\;,
\end{equation}
the cancelling minus sign arising because the light is approaching us.
That is,
\begin{equation}
\chi=\ln\left({t_o\over t_e}\right)\;.
\end{equation}
Thus, according to Equation~(4), the cosmological redshift is
\begin{eqnarray}
\hskip 2.4in 1+z&=&{a(t_o)\over a(t_e)}={t_o\over t_e}\nonumber\\
&=& e^\chi\;,
\end{eqnarray}
fully consistent with the result we derived in Equation~(17) through
a consideration of the time dilation between moving frames (Equation~11)
and its subsequent modification as a result of the shift in arrival
times (Equation~13). So in the Milne cosmology, the redshift may be 
calculated either from knowledge of the expansion factor $a(t)$, 
or by using a more direct approach already understood in the 
context of general relativity that does not involve the assumption
of an expanding space.

\subsection{de Sitter Space}
The de Sitter spacetime is the first of the six static FRW solutions we will
encounter that has a constant, but nonzero, curvature. Unlike
the Milne Universe, which describes a flat universe with no gravitational 
acceleration, de Sitter has $\rho\not=0$, and objects
not only recede from one another, but also accelerate under the influence
of gravity. Since the spacetime in de Sitter is curved, this FRW
metric provides us with an important validation of our method,
complementary to the Milne example.

The de Sitter cosmology (de Sitter 1917) corresponds to a universe devoid
of matter and radiation, but filled with a cosmological constant whose
principal property is the equation of state $p=-\rho$. The FRW metric in
this case may be written
\begin{equation}
ds^2 = c^2 dt^2-e^{2Ht} [dr^2 + r^2d\Omega^2]\;,
\end{equation}
where $k = 0$ and the expansion factor has the specific form
\begin{equation}
a(t)=e^{Ht}\;,
\end{equation}
in terms of the Hubble constant $H$. This cosmology may represent
the Universe's terminal state, and may also have corresponded
to its early inflationary phase, where it would have produced an
exponentiation in size due to the expansion factor $\exp(Ht)$.

Unlike the Minkowski and Milne models, the de Sitter cosmology contains mass-energy
(in the form of a cosmological constant). However, an observer using
only comoving coordinates is in free fall and is unaware of the
gravitational acceleration. This was Einstein's ``happiest thought
of his life" that lead to the Principle of Equivalence, which states
that the spacetime in a free falling frame is locally Minkowskian,
consistent with special relativity. But we realize that gravity plays
an important role when we instead move to a different set of
coordinates (Melia 2007; Melia \& Abdelqader 2009), 
which may include the proper radius $\eta(t)=a(t)r$.

Let us first present the transformation that casts this metric into
its static form, and then discuss the physical meaning of the new
coordinates. With the transformation
\begin{eqnarray}
\hskip2.5in \eta&\hskip-0.1in=\hskip-0.1in&a(t)r\nonumber\\
T&\hskip-0.1in=\hskip-0.1in&t-{1\over 2H}\ln\Phi\;,
\end{eqnarray}
where
\begin{equation}
\Phi\equiv 1-\left({\eta\over R_{\rm h}}\right)^2\;,
\end{equation}
and
\begin{equation}
R_{\rm h}\equiv {c\over H}
\end{equation}
is the gravitational (or Hubble) radius, the de Sitter metric becomes
\begin{equation}
ds^2=\Phi\, c^2\,dT^2-\Phi^{-1}d\eta^2 - \eta^2d\Omega^2\;.
\end{equation}
Clearly, $g_{TT}=\Phi$ and $g_{\eta\eta}=\Phi^{-1}$, both
independent of $T$.

Written in this way, the metric explicitly reveals the spacetime
curvature most elegantly inferred from the corollary to Birkhoff's
theorem (Birkhoff 1923). This theorem states that in a spherically
symmetric spacetime, the only solution to the Einstein equations
is the Schwarzschild exterior solution, which is static. What is
relevant to our discussion here is not so much the Birkhoff theorem
itself, but rather its very important corollary. The latter is a
generalization of a well-known result of Newtonian theory, that
the gravitational field of a spherical shell vanishes inside
the shell. The corollary to Birkhoff's theorem states that the metric
inside an empty spherical cavity, at the center of a spherically
symmetric system, must be equivalent to the flat-space Minkowski
metric. Space must be flat in a spherical cavity even if the system is
infinite. It matters not what the constituents of the medium outside
the cavity are, as long as the medium is spherically symmetric.

If one then imagines placing a spherically symmetric mass at
the center of this cavity, according to Birkhoff's theorem and its
corollary, the metric between this mass and the edge of the cavity
is necessarily of the Schwarzschild type. Thus, the worldlines
linked to an observer in this region are curved relative to the center
of the cavity in a manner determined solely by the mass we have
placed there. This situation may appear to contradict our assumption
of isotropy, which one might naively take to mean that the spacetime
curvature within the medium should cancel since the observer sees
mass-energy equally distributed in all directions. In fact, the
observer's worldlines are curved in every direction because,
according to the corollary to Birkhoff's theorem, only the mass energy
between any given pair of points in this medium affects the path
linking those points.

The form of the metric in Equation~(26) is how de Sitter himself
first presented his now famous solution. One can almost see the
inspiration for it by considering Schwarzschild's solution
describing the spacetime around an enclosed, spherically symmetric
object of mass $M$:
\begin{equation}
ds^2 = c^2\,dT^2[1-2GM/c^2\eta]-d\eta^2[1-2GM/c^2\eta]^{-1}-\eta^2d\Omega^2\;.
\end{equation}
De Sitter's metric describes the spacetime around a radially dependent
enclosed mass $M(\eta)$. In a medium with uniform mass-energy density,
\begin{equation} 
M(\eta) = M(R_{\rm h})(\eta/R_{\rm h})^3\;,
\end{equation}
for which the Schwarzschild factor $1-(2GM/c^2\eta)$ transitions into
$1-(\eta/R_{\rm h})^2$, what we have here called $\Phi$ (see Equation~24).
It should be emphasized that this Equation implicitly
contains the restriction that no mass energy beyond $\eta$ should contribute
to the gravitational acceleration inside of this radius, as required by the
corollary to Birkhoff's theorem. For a given interval $ds$, the
time $T$ clearly diverges as $\eta$ approaches $R_{\rm h}$, which therefore
represents the limiting distance beyond which the spacetime curvature
prevents any signal from ever reaching us. Though we know it as the
Hubble radius, $R_{\rm h}$ is actually defined as a Schwarzschild radius,
by the condition
\begin{equation}
{2GM(R_{\rm h})\over c^2}=R_{\rm h}\;.
\end{equation}
That is, $R_{\rm h}$ is in fact the distance at which the enclosed mass-energy
is sufficient to turn it into the Schwarzschild radius for an observer at the
origin of the coordinates. And it is trivial to show that for $k=0$, $R_{\rm h}$
reduces to its more recognizable Hubble manifestation in Equation~(25).

The point of all this is for us to recognize that de Sitter's metric in Equation~(26)
is not only static (as we require for our procedure), but that it also
contains the effects of gravitational curvature through the factor $\Phi$.
We will now follow steps 2 and 3 in our procedure, as we did with Milne, to
derive the cosmological redshift in de Sitter based on the Doppler and
gravitational effects.

The time dilation is here given as
\begin{equation}
{dt\over dT}=\left[\Phi-{1\over c^2}\Phi^{-1}\left({d\eta\over dT}\right)^2\right]^{1/2}\;,
\end{equation}
again assuming that the source moves only with the Hubble flow. Since $r$ is
therefore constant, we also have
\begin{equation}
{d\eta\over dT}={\partial\eta\over \partial t}{dt\over dT}=\dot{a}r{dt\over dT}
=H\eta\,{dt\over dT}\;.
\end{equation}
Equations~(23), (30), and (31) are therefore consistent with
\begin{equation}
{dt\over dT}=\Phi\;.
\end{equation}
This time dilation includes {\it both} the effects of gravity and the kinematics
associated with the Hubble recession of the source. But as we learned
previously, we cannot yet use this to infer the shift in frequency of
the light without first finding the {\it apparent} time dilation, analogous
to Equation~(13).

In de Sitter, the proper velocity component of the source along our line-of-sight is
\begin{equation}
v_\eta\equiv\sqrt{g_{\eta\eta}\over g_{TT}}{d\eta\over dT}\;,
\end{equation}
where now neither $g_{\eta\eta}$ nor $g_{TT}$ are equal to $1$.
Thus, all told
\begin{eqnarray}
\hskip2.5in{\nu_o\over\nu_e}&\hskip-0.1in=\hskip-0.1in&\left(1+{v_\eta\over c}\right)^{-1}{dt\over dT}
\,{\bigg{|}}_{T_e}\nonumber\\
&\hskip-0.1in=\hskip-0.1in&\left(1+{\eta(T_e)\over R_{\rm h}}\right)^{-1}\Phi(T_e)\;,
\end{eqnarray}
which means that in de Sitter
\begin{equation}
1+z\equiv {\nu_e\over \nu_o}=\left[1-{\eta(T_e)\over R_{\rm h}}\right]^{-1}\;.
\end{equation}
According to Equation~(4), this expression should be equivalent to
$a(t_0)/a(t_e)$, so let us see if this is indeed the case.

Along a null geodesic from $t_e$ to $t_o$, we have
\begin{equation}
\int_0^r dr'=c\int_{t_e}^{t_o} {dt'\over \exp{(Ht')}}\;,
\end{equation}
so
\begin{equation}
r={c\over H}\left(e^{-Ht_e}-e^{-Ht_o}\right)\;.
\end{equation}
That is,
\begin{equation}
\eta(t_e)=a(t_e)r={c\over H}\left(1-e^{-H(t_o-t_e)}\right)\;.
\end{equation}
And therefore
\begin{equation}
1+z={a(t_o)\over a(t_e)}=e^{H(t_o-t_e)}=
\left[1-{\eta(T_e)\over R_{\rm h}}\right]^{-1}\;,
\end{equation}
fully consistent with the result in Equation~(35). As we found in the
case of Milne, the cosmological redshift in de Sitter may be calculated
either from the expansion factor $a(t)$, or from the time dilation and 
frequency shift associated with motion of the source. For several reasons, 
the de Sitter case is even more important than Milne in this discussion because it clearly
represents a situation in which the redshift is due to {\it both} gravitational
and kinematic effects. We see in both cosmologies that the interpretation
of redshift as an expansion of space is dependent upon the coordinates one
chooses to calculate $z$.

\subsection{The Lanczos Universe}
The Lanczos Universe (Lanczos 1924) is described by the metric
\begin{equation}
ds^2=c^2dt^2-\left({cb}\right)^2\cosh^2(t/b)\left[{dr^2\over 1-r^2}+
r^2d\Omega^2\right]\;,
\end{equation}
where $b$ is a constant (though not the Hubble constant $H\equiv\dot{a}/a$) 
and $k=+1$. The expansion 
factor is $a(t)=(cb)\cosh(t/b)$, so $H=(1/b)\tanh(t/b)$. This solution represents
the gravitational field of a rigidly rotating dust cylinder coupled to a
cosmological constant. We use the following transformation (see Florides 1980)
to render this metric in static form:
\begin{equation}
\eta={cbr}\cosh(t/b)\;,
\end{equation}
and
\begin{equation}
\tanh(T/b)=\left(1-r^2\right)^{-1/2}\tanh(t/b)\;,
\end{equation}
which together allow us to write the interval in the form
\begin{equation}
ds^2=\left[1-\left({\eta\over cb}\right)^2\right]c^2dT^2-
\left[1-\left({\eta\over cb}\right)^2\right]^{-1}d\eta^2-\eta^2d\Omega^2\;.
\end{equation}

We now follow the steps used for the Minkowski, Milne, and de Sitter metrics, and first
calculate the time dilation
\begin{equation}
{dt\over dT}=\left[\left(1-\left({\eta\over cb}\right)^2\right)-{1\over c^2}
\left(1-\left({\eta\over cb}\right)^2\right)^{-1}\left({d\eta\over dT}\right)^2\right]^{1/2}\;.
\end{equation}
But since $dr=0$ (and therefore $dr/dT=0$) in the Hubble flow, we have
\begin{equation}
{d\eta\over dT}=cr\sinh(t/b){dt\over dT}\;.
\end{equation}
From Equations~(44) and (45) we
therefore see that
\begin{equation}
{dt\over dT}={1-r^2\cosh^2\left(t/b\right)\over \sqrt{1-r^2}}
\end{equation}
(which may also be confirmed directly from Equation~42).
Thus, following the same argument as before, the ratio of the frequency of light
actually measured by the observer to that emitted is given by Equation~(13), where now
\begin{equation}
v_\eta\equiv \sqrt{g_{\eta\eta}\over g_{TT}}{d\eta\over dT}={cr\sinh\left(t/b\right)\over
\sqrt{1-r^2}}\;.
\end{equation}
Therefore
\begin{equation}
{\nu_0\over \nu_e}={1-r^2\cosh^2(t/b)\over\sqrt{1-r^2}+r\sinh(t/b)}\,{\bigg{|}}_{T_e}\;,
\end{equation}
and the redshift in this cosmology is given by
\begin{equation}
1+z={\nu_e\over \nu_0}={\sqrt{1-r^2}+r\sinh(t/b)\over 1-r^2\cosh^2(t/b)}\,{\bigg{|}}_{T_e}\;.
\end{equation}

To compare this expression with the result we would have obtained from Equation~(4), consider
the propagation of a light signal from its emission at comoving distance $r_e$ at time $t_e$, on its way
to the observer at $r=0$ and time $t_0$. The geodesic equation describing this trajectory
(derived from Equation~40) is
\begin{equation}
\int_{0}^{r_e}{dr\over\sqrt{1-r^2}}=\int_{t_e/b}^{t_0/b}{du\over \cosh(u)}\;,
\end{equation}
whose solution may be written
\begin{equation}
\sin^{-1}\left(r_e\right)=2\tan^{-1}\left(e^{t_0/b}\right)-2\tan^{-1}\left(e^{t_e/b}\right)\;.
\end{equation}
Therefore,
\begin{equation}
r_e=2\sin\left[\tan^{-1}\left(e^{t_0/b}\right)-\tan^{-1}\left(e^{t_e/b}\right)\right]
\cos\left[\tan^{-1}\left(e^{t_0/b}\right)-\tan^{-1}\left(e^{t_e/b}\right)\right]\;,
\end{equation}
and after some algebra,  using the identities $\sin\left(\tan^{-1}\left[x\right]\right)=x\left(1+x^2\right)^{-1/2}$ and
$\cos\left(\tan^{-1}\left[x\right]\right)=\left(1+x^2\right)^{-1/2}$, one finds that
\begin{equation}
r_e=2\left(e^{t_0/b}-e^{t_e/b}\right)\left(1+e^{(t_0+t_e)/b}\right)
\left(1+e^{2t_0/b}\right)^{-1}\left(1+e^{2t_e/b}\right)^{-1}\;.
\end{equation}
With further lengthy algebraic manipulations, substituting this expression
into Equation~(49) produces the final result,
\begin{equation}
1+z={\cosh\left(t_0/b\right)\over\cosh\left(t_e/b\right)}\;,
\end{equation}
which is the correct form of Equation~(4) for the Lanczos expansion factor
$a(t)=(cb)\cosh(t/b)$.

\subsection{A Lanczos Universe with $k=-1$}
The application of our procedure to the next case is very similar to Lanczos, so there is no 
need to dwell on the various steps. The fifth static FRW metric is simply the Lanczos 
Universe with $k=-1$, for which
\begin{equation}
ds^2=c^2dt^2-\left({cb}\right)^2\sinh^2(t/b)\left[{dr^2\over 1+r^2}+
r^2d\Omega^2\right]\;,
\end{equation}
where $a(t)=(cb)\sinh(t/b)$. The metric may be written in static form with the
transformation
\begin{equation}
\eta={cbr}\sinh(t/b)\;,
\end{equation}
and
\begin{equation}
\tanh(T/b)=\left(1+r^2\right)^{1/2}\tanh(t/b)\;,
\end{equation}
which together allow us to write the interval in the form
\begin{equation}
ds^2=\left[1-\left({\eta\over cb}\right)^2\right]c^2dT^2-
\left[1-\left({\eta\over cb}\right)^2\right]^{-1}d\eta^2-\eta^2d\Omega^2\;,
\end{equation}
identical (in terms of $\eta$ and $T$) to the Lanczos metric in Equation~(43). 
We see immediately that the redshift in this case is also given
by Equation~(44) though, of course, $\eta$ and $T$ are here given by 
Equations~(56) and (57), respectively, instead of (41) and (42). 
Therefore, in this case we have
\begin{equation}
{d\eta\over dT}=cr\cosh(t/b){dt\over dT}\;,
\end{equation}
and
\begin{equation}
{dt\over dT}={1-r^2\sinh^2\left(t/b\right)\over \sqrt{1+r^2}}
\end{equation}
(which may also be confirmed directly from Equation~57).

The proper velocity is thus
\begin{equation}
v_\eta\equiv \sqrt{g_{\eta\eta}\over g_{TT}}{d\eta\over dT}={cr\cosh\left(t/b\right)\over
\sqrt{1+r^2}}\;,
\end{equation}
and the redshift analogous to Equation~(49) is
\begin{equation}
1+z={\nu_e\over \nu_0}={\sqrt{1+r^2}+r\cosh(t/b)\over 1-r^2\sinh^2(t/b)}\,{\bigg{|}}_{T_e}\;.
\end{equation}
The exercise is completed by calculating $r_e\equiv r(t_e)$ from the geodesic equation
\begin{equation}
\int_0^{r_e}{dr\over\sqrt{1+r^2}}=\int_{t_e/b}^{t_0/b}{du\over \sinh(u)}\;,
\end{equation}
whose solution is
\begin{equation}
\sinh^{-1}\left(r_e\right)=\ln\left(\tanh\left[t_0/2b\right]\right)-\ln\left(\tanh\left[t_e/2b\right]\right)\;.
\end{equation}
That is,
\begin{equation}
r_e={1\over 2}\left({\tanh(t_0/2b)\over \tanh(t_e/2b)}
-{\tanh(t_e/2b)\over \tanh(t_0/2b)}\right)\;.
\end{equation}
Another lengthy algebraic manipulation following the substitution of this expression into Equation~(62)
produces the result
\begin{equation}
1+z={\sinh\left(t_0/b\right)\over\sinh\left(r_e/b\right)}\;,
\end{equation}
which matches  the correct form of Equation~(4) for the expansion factor appropriate for this metric.

\subsection{Anti-de Sitter Space (A Universe with Negative Mass Density)}
The sixth, and final, static FRW metric is that for a Universe with negative mass density
and spatial curvature $k=-1$. Known as anti-de Sitter space, due to its negative spacetime
curvature, this metric is given by
\begin{equation}
ds^2=c^2dt^2-\left({cb}\right)^2\sin^2\left(t/b\right)\left[{dr^2\over 1+r^2}+
r^2d\Omega^2\right]\;,
\end{equation}
where clearly the expansion factor is now $a(t)={cb}\sin\left(t/b\right)$.
The coordinate transformation
\begin{equation}
\eta={cbr}\sin(t/b)\;,
\end{equation}
and
\begin{equation}
\tan(T/b)=\left(1+r^2\right)^{1/2}\tan(t/b)\;,
\end{equation}
produces the static form of the metric,
\begin{equation}
ds^2=\left[1+\left({\eta\over cb}\right)^2\right]c^2dT^2-
\left[1+\left({\eta\over cb}\right)^2\right]^{-1}d\eta^2-\eta^2d\Omega^2\;.
\end{equation}
The proper velocity is now
\begin{equation}
v_\eta={cr\cos(t/b)\over\sqrt{1+r^2}}\;,
\end{equation}
where
\begin{equation}
{dt\over dT}={1+r^2\sin^2\left(t/b\right)\over \sqrt{1+r^2}}\;.
\end{equation}
For this metric, the redshift is therefore
\begin{equation}
1+z={\nu_e\over \nu_0}={\sqrt{1+r^2}+r\cos(t/b)\over 1+r^2\sin^2(t/b)}\,{\bigg{|}}_{T_e}\;.
\end{equation}

Now, along a geodesic, 
\begin{equation}
\int_0^{r_e}{dr\over\sqrt{1+r^2}}=\int_{t_e/b}^{t_0/b}{du\over \sin(u)}\;,
\end{equation}
which has the solution
\begin{equation}
r_e={1\over 2}\left({\tan(t_0/2b)\over \tan(t_e/2b)}
-{\tan(t_e/2b)\over \tan(t_0/2b)}\right)\;.
\end{equation}
And substituting this expression for $r_e$ into Equation~(73) then gives
\begin{equation}
1+z={\sin(t_0/b)\over\sin(t_e/b)}\;,
\end{equation}
which is again the correct form of the redshift in terms of the expansion factor for
this metric.

\section{Conclusions}
The cosmological redshift has the same form in terms of the expansion factor
regardless of whether the spacetime curvature is constant or not. In this paper,
we have focused on the six static FRW metrics, and showed for them that the 
interpretation of $z$ as due to the ``stretching of light" in an expanding space 
is coordinate dependent. When calculated using an alternative set of coordinates,
the redshift has precisely the same contributions---Doppler and gravitational 
shifts---that one 
would expect from the calculation of the lapse function in other applications
of general relativity. This association may break down for the non-static
FRW metrics, but it would be difficult to see why, given that the formulation of
$z$ in terms of the expansion factor $a(t)$ is identical in all cases. Still, the
proof we have presented here is only partial. It remains to be seen whether
the cosmological redshift is a lapse function even when the spacetime
curvature changes with time.

There are many reasons why the distinction between an expanding space
and a fixed space through which particles move is dynamically important.
For example, one sometimes hears statements to the effect that light in cosmology 
can be transported over vast distances faster than one would infer on
the basis of $c$ alone. The justification for this is that the speed of light
is limited to $c$ only in an inertial frame, but if space is expanding, then
light can be carried along with the expansion at even higher speeds.
However, it is not difficult to understand why such notions arise from 
the improper use of the coordinates. In general relativity, the velocity
measured by an observer is the proper velocity (e.g., Equation~12), 
calculated in terms of the proper distance and proper time. For light,
$ds$ satisfies the null condition (i.e., $ds=0$) and therefore
$v_\eta$ is {\it always} equal to $c$, regardless of which coordinate
system is being used, or even if the frame of reference is inertial or not.
What {\it is} true is that the speed $d\eta/dt$ is not restricted to $c$.
But this is not the proper speed measured by a single observer using
solely his rulers and clocks, because the quantity $\eta=a(t)r$ is
a community distance, compiled from the infinitesimal contributions
of myriads of observers lined up between the endpoints (see, e.g.,
Weinberg 1972).  A demonstration that $z$ is not due to the stretching
of space affirms these conclusions by removing the possibility that
light may be ``carried along" superluminally with the expansion.

Though we have only partially addressed the issue concerning
the origin of cosmological redshift, we can now nonetheless turn 
these results around and ask the opposite question. If there really
exists a third mechanism producing a redshift, beyond Doppler
and gravity, why don't we see it manifestated in the static
FRW metrics?  After all, FRW spacetimes with constant curvature
also satisfy Equation~(4), just like the rest do. And if Equation~(4)
is evidence that $z$ arises from the stretching of light in an expanding
space, this process should happen regardless of whether the metric 
is static or not.

In closely related work, Chodorowski (2011) uses a very different technique
to arrive at results similar to those reported in this paper. The fact that these
two approaches lead to the same conclusions adds significantly to the validity
of (his and) our thesis that cosmological redshift is not due to a new form of 
wavelength extension, beyond those from kinematic and gravitational effects. 
Chodorowski's approach is beautifully complementary to that described
here because very different coordinates systems are utilized in the
derivations. We have sought metrics that can be written in
static form, though the transformed coordinates do not necessarily
describe a local inertial frame. Yet the velocity of the source may be
expressible in terms of these coordinates, as long as we correctly
use the proper distance and proper time to evaluate this (proper)
velocity. (By the way, this is what we typically do with the 
Schwarzschild and Kerr metrics.) The decomposition of the cosmological
redshift is then based on this proper velocity. If the velocity is zero
in this coordinate system, then the time dilation is entirely due to
the curvature (or gravity), but the cosmological redshift generally
includes a second factor that enters because sources are moving
with the Hubble flow. Chodorowski instead chooses to parallel-transport
the source's velocity into the local inertial frame of the central observer,
thereby providing a means of calculating the ``Dopplerian" redshift
(as he calls it) in this frame, with ``the rest" arising from the effects
of curvature. What's interesting, of course, is that because the
two sets of coordinates are different (one inertial, the other not),
the two decompositions are generally not equal, but the final
results are the same, as they should be because the underlying
physics is identical. 

Demonstrating that $z$ is a lapse function even for the time-dependent
FRW metrics is quite challenging. But given the importance of understanding 
the origin of cosmological redshift, it is a task worth undertaking. We
mention in this regard that Mizony \& Lachi\`eze-Rey (2005) found a
way of transforming an FRW metric into a local static form, which
interestingly is equivalent to de Sitter in this limited domain. Following
their approach may be a very useful intermediate step in the process
of finding the lapse function globally in cases where the spacetime
curvature is not constant. We will examine this question next 
and hope to report the results of these efforts in the near future.

\section*{Acknowledgments}
This research was partially supported by ONR grant N00014-09-C-0032
at the University of Arizona, and by a Miegunyah Fellowship at the
University of Melbourne. I am particularly grateful to Amherst College
for its support through a John Woodruff Simpson Lectureship. Finally,
I wish to thank Roy Kerr and Andrew Shevchuk for many helpful and
enjoyable discussions, and the anonymous referee for bringing 
Chodorowski's recent paper to my attention and for providing other
helpful comments.

\end{document}